\documentclass[preprint,amssymb,showpacs,supersriptaddress,floatfix]{revtex4-2}
\usepackage{graphicx}
\usepackage{dcolumn}
\usepackage{bm}
\usepackage{color}

\usepackage[utf8]{inputenc}
\usepackage[T1]{fontenc}
\usepackage{mathptmx}
\usepackage{etoolbox}

\begin{document}

\preprint{AIP/123-QED}


\title {Effect of  memristor´s potentiation-depression curves peculiarities in the convergence of physical perceptrons}


\author{Walter Quiñonez$^{1}$, María José Sánchez$^{1,2}$, Diego Rubi$^{1}$}

\email{Corresponding author. Electronic mail: diego.rubi@gmail.com}

\affiliation{$^{1}$ Instituto de Nanociencia y Nanotecnología (INN), CONICET-CNEA, Buenos Aires and Bariloche, Argentina \\ $^{2}$ Centro Atómico Bariloche and Instituto Balseiro (Universidad Nacional de Cuyo), 8400 San Carlos de Bariloche, Río Negro, Argentina }

\date{\today}

\begin{abstract}

Neuromorphic computing aims to emulate the architecture and information processing mechanisms of the mammalian brain. This includes the implementation by hardware of neural networks. Oxide-based memristor arrays with cross-bar architecture appear as a possible physical implementation of neural networks. 

{In this paper, we obtain experimental potentiation-depression (P-D) curves on different manganite-based memristive systems and simulate the learning process of perceptrons for character recognition. We analyze how the specific characteristics of the P-D curves affect the convergence time -characterized by the EPOCHs-to-convergence (ETC) parameter- of the network. Our work shows that ETC is reduced for systems displaying P-D curves with relatively low granularity and non-linear and asymmetric response. In addition, we also show that noise injection during the synaptic weight actualization further reduces the ETC. The results obtained here are expected to contribute to the optimization of hardware neural networks based on memristors cross-bar arrays.}

\vspace{10px}


\end{abstract}

\maketitle

\section{Introduction}

Neural networks (NN) are one of the key algorithms in the broad field of Machine Learning (ML) \cite{Geron}, applied nowadays to tackle an enormous variety of problems such as object detection and recognition, natural language processing, healthcare and medical applications or stock market predictions, among others \cite{sarker_2021}. NN are usually implemented by software running on standard computers with the Von Neumann architecture, where the units that process and store the information are physically separated. This architecture faces the so-called "memory wall" or "Von Neumann bottleneck", originated in the intense data traffic between memory and processing units through a bus in between working at a lower frequency  than the processor, which significantly enlarges the computing time. In addition, the Von Neumann architecture is highly energetically inefficient. For instance, the training of a state-of-the-art NN requires thousands of processing units consuming around 200 W each \cite{Mehonic2022}. Given the exponential increase of NN used in our Society of Information, their carbon footprint and energetic sustainability is becoming a significant concern. This motivates the design and construction of computers beyond the Von Neumann paradigm, able to better deal with the parallel processing of large amounts of data -as required by NN- with significantly lower energy consumption.

Unlike modern computers, the mammalian brain is able to perform complex tasks such as speech or image recognition with an energy consumption of only 20 W. In addition, the brain is significantly error tolerant and able to deal with noisy signals. The main difference between the brain and Von Neumann computers is in their architecture; the former being composed by a intricate network of $10^{11}$ neurons connected through $10^{15}$ synapses. The learning capability of the brain is usually attributed to the (analog) adaptability of synapses, which are also responsible of information storage. A possible strategy to develop more efficient computers is by replicating the architecture and information processing ways of the brain. This has been coined as "neuromorphic computing" (NC) and has been a subject of intense research in the last years \cite{yu_2017,Mehonic2022}. 

Among solid-state devices with high potential for NC are the so-called "memristors", which are essentially metal-insulator-metal micro or nanostructures able to change their electrical resistance between different states upon the application of electrical stimuli \cite{iel_2016,saw_2008}. This change could be either non-volatile or volatile, and several synaptic/neuron functionalities with different complexity can be therefore electrically replicated 
\cite{Wu_2022,Kumar2022}. Memristive effects have been found for an enormous variety of materials, including simple \cite{Strukov_2008} and complex \cite{Liu_2000} oxides, ferroelectrics \cite{chan_2012}, topotactic redox perovskites \cite{Acevedo_2020, roman_2022}, halides \cite{Fang_2021} or 2D materials such as graphene \cite{Schranghamer_2020}, among others. Memristive effects in oxide based compounds usually rely on the electromigration of charged defects, such as oxygen vacancies \cite{saw_2008}. 

The development of neuromorphic circuits based on memristors requires their integration in multiple device architectures. A possible implementation are the cross-bar arrays, which consist in M bit and N word metallic lines (the latter oriented 90 deg with respect to the former), sandwiching the active memristive oxide. This gives an array of M x N memristors which can be individually accessed through the bit/word lines.

If voltages $V_j$ are applied to the bit lines (input), Ohm and Kirchhoff laws indicate that currents $I_i$ should be produced in the word lines (output) according to:

\begin{equation}
    I_i = \sum_{j=1}^{M} G_{ij} V_j,
\label{eq_current}
\end{equation}
where $G_{ij}$  is the electrical conductance of the memristor  located at the site ${ij}$ in the array. This is essentially a matrix-vector product performed by hardware in a single step. It has been shown that memristor cross-bar arrays can be implemented for in-memory computing processes, such as the resolution of linear algebra problems \cite{Zhong_2019,Zhong_2020}.

Memristor cross-bar arrays have also been implemented as physical NN, starting with the seminal work of Strukov et al. where they used TiO$_2$-based memristors for the developing of simple perceptrons for character recognition \cite{Alibart_2013,Prezioso_2014}. Memristor conductances are linked to the NN synaptic weights, which are progressively adapted during the learning process by using the memristor potentiation (P)-depression (D) properties. Different NN have been physically implemented afterwards \cite{Chen_2015,Li_2018}, including convolutional NN \cite{Joshi_2020, Jang_2022} and resevoir computing systems \cite{Zhu_2020}. Hebbian learning based on spike-time-dependent plasticity \cite{kumar_2023} and 4-bit multilevel operation \cite{kim_2022} have been also reported for memristor cross-bar arrays.

A critical issue related to the development of physical NN are the so-called "non-idealities" of memristors, which can affect the NN performance \cite{Chen_2015, Li_2018, Shibata_2020, Gutsche_2021, Bengel_2021, Yon_2022, Apsangi_2022}. These include distinct P-D curves with constrained conductance windows $\Delta G$ and discrete number of conductance levels N (the granularity of the curves can be defined as the ratio between the latter and the former). Other non-idealities include the intrinsic stochasticity of the memristive response -which impacts both on the writing and reading of their conductive states-, device-to-device and cycle-to-cycle variations or stuck devices. It is commonly accepted that linear/symmetric P-D curves are preferred as they can improve the accuracy of the network \cite{Chen_2015, Burr_2015, Shibata_2020,Apsangi_2022}. However, some controversy remains on this point as recent reports indicate that the impact of the P-D linearity on the NN performance might be dependent on specific details of the training protocol, such as the synaptic weights adaptation procedure \cite{Gutsche_2021}, evidencing the need of further systematic studies to better understand this issue.

On the other hand, it is well established on standard NN that the addition of noise during the training can improve the NN generalization ability and reduce training losses \cite{Yon_2022}. A usual interpretation for the latter is that a reasonable amount of noise helps the system to avoid the Loss function to stabilize on local minima, favouring the convergence to a global minimum. The noise can be injected in different stages: in the train data set -strategies such as data augmentation are broadly used for this purpose-, on the synaptic weight actualization or on the activation function \cite{Joshi_2020}. For the case of memristor-based physical NN, the addition of noise has been shown helpful to maintain the NN accuracy in the presence of memristor non-idealities \cite{Joshi_2020} and also to avoid over-fitting \cite{Du_2022}.

In this work, we experimentally determine the P-D response of different manganite-based memristive devices and use these data to simulate the learning process of simple perceptrons based on memristor cross-bar arrays. We found differences in the network performance for each system, namely on the number of iterations (EPOCHs) needed to achieve convergence. With the help of synthetic P-D curves, which allow varying non-idealities in a one-by-one basis, we determined the influence of the P-D curves granularity and non-linearity/asymmetry on the EPOCHs-to-convergence (EPC). Further, we found that EPC can be substantially reduced by adding noise to the weight actualization during the training procedure. Our results are relevant as they allow defining strategies for minimizing the EPC, which might be useful for systems that need continuous retraining, as it is the case of memristor-based NN with progressive conductance drifts of individual devices.

\section{Experimental manganite-metal systems}

We have characterized the P-D response of three memristive manganite-based single devices: i) Pt/La$_{0.33}$Ca$_{0.67}$MnO$_3$/Ag (LCMO-Ag), ii) Pt/La$_{0.33}$Ca$_{0.67}$MnO$_3$/Ti (LCMO-Ti) and iii) Nb:SrTiO$_3/$La$_{0.5}$Sr$_{0.5}$Mn$_{0.5}$Co$_{0.5}$O$_{3-x}$/Pt (NSTO-LSMCO). In the case of LCMO-Ag and LCMO-Ti devices, thin polycrystalline LCMO layers ($\approx$ 100 nm thick) were grown by laser ablation on commercial platinized silicon wafers. Top metallic electrodes were defined by electronic lithography and consisted in Ag or Ti micropillars (25 and 100 $\mu m^2$, respectively) embedded in an insulating SiO$_2$ matrix. The micropillars end in millimeter sized pads suitable to make electrical contact. Ti was covered with a thin Au layer to avoid its oxidation upon exposure to ambient conditions. More details about the fabrication process and device geometry can be found in Refs. \cite{Miranda_2017,acevedo_2018}. For the NSTO-LSMCO device, epitaxial LSMCO thin films ($\approx$ 20 nm thick) were grown by laser ablation on Nb-doped (001) SrTiO$_3$ single crystals. Top Pt electrodes (4 x$10^4$ $\mu m^2$) were defined by optical lithography. The NSTO substrate was used as bottom electrode.

Electrical characterization was performed at room temperature with a source measure unit Keithley 2612B hooked to a commercial probe station. Bottom electrodes were grounded in all cases and the stimulation was applied to the top electrode. 

\begin{figure}
  \centering    
  \includegraphics[scale=0.45]{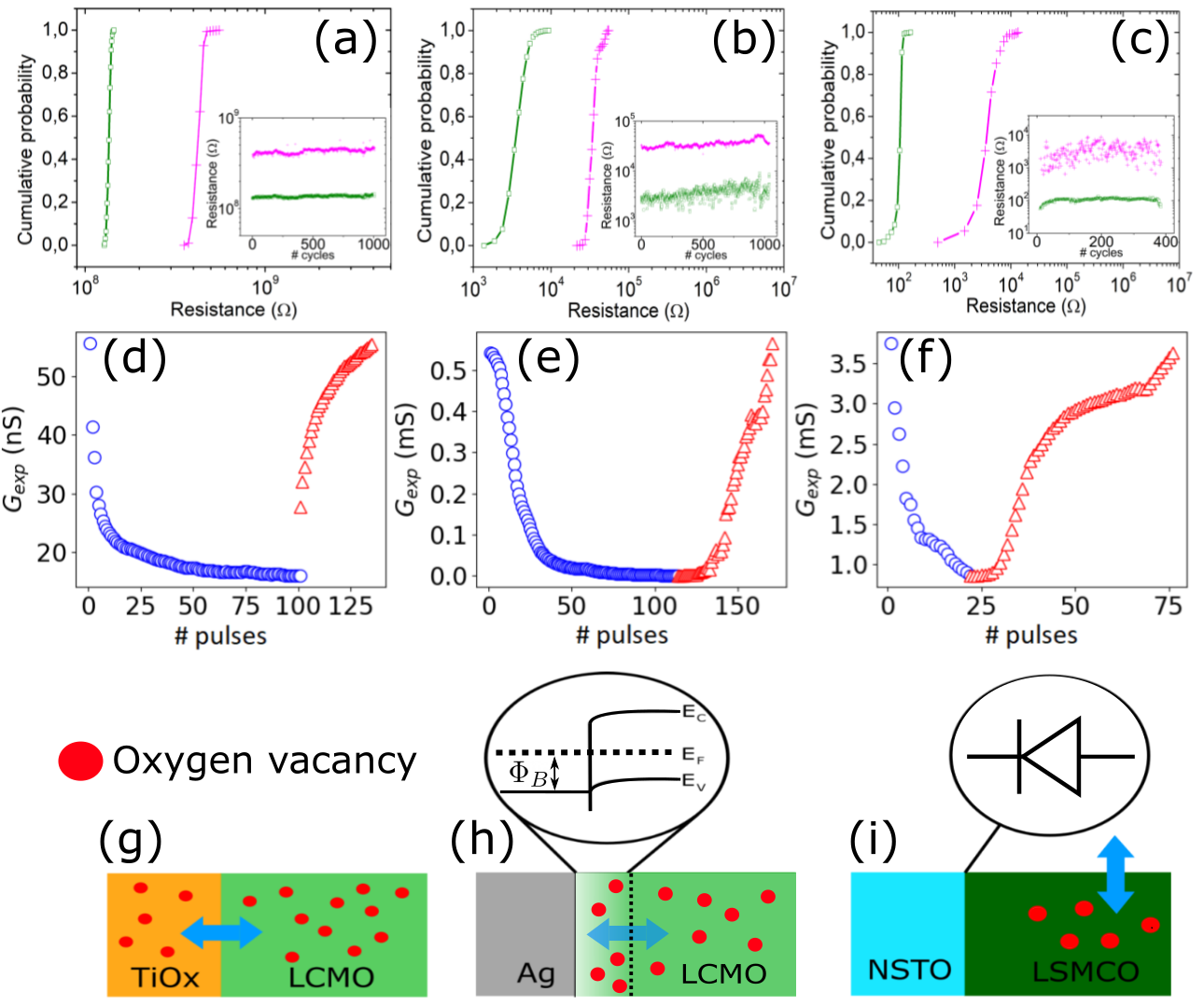}
  \caption{\textit{(a), (b), (c) Endurance tests (insets) and cumulative probabilities (main panels) for LCMO/Ti, LSMCO/Ag and NSTO/LSMCO devices, respectively. The SET process is achieved by applying -2.5 V, -2 V and -4 V, respectively, while the RESET is obtained after the application of +2 V, +1.5 V and +7 V, respectively. In all cases the pulse time-width was around 1 ms. (d), (e), (f) Experimental potentiation (P, red symbols) and depression (D, blue symbols) curves measured on LCMO/Ti, LCMO/Ag and NSTO/LSMCO devices, respectively. See the main text for further details. (g), (h), (i) Sketches depicting the memristive mechanisms for LCMO/Ti, LCMO/Ag and NSTO/LSMCO devices, respectively. See the main text for further details}}
\label{Fig1}
\end{figure}

Figures \ref{Fig1}(a), (b), (c) (insets) display resistance switching between high and low states, upon the application of single pulses, for LCMO/Ti (1000 cycles), LCMO/Ag (1000 cycles) and NSTO-LSMCO (400 cycles) devices, respectively. The main panels of the same figures display the calculated cumulative probabilities, where well defined ON-OFF windows are seen in all cases.

We notice that  the  memristive response of each device is ruled by a specific mechanism/process. For the LCMO/Ti device, due to the reactive nature of Ti, an ultrathin TiO$_x$ layer is formed upon Ti deposition on LCMO. The memristive behavior is driven by a redox reaction between LCMO and TiO$_x$ layers. The former (later) is a p-type (n-type) oxide and thus its resistance  increases (decreases)  as the oxygen vacancies (OV) content  is increased (decreased). When oxygen is transferred from LCMO to TiO$_x$, both layers increase their resistance (RESET process); in the opposite case, both layers decrease their resistance and the SET process takes place \cite{acevedo_2018, fer_2019}. 

For the LCMO/Ag device, the electronic transport takes place along multiple conducting pathways spanning the oxide layer with their resistance dominated by the potential barrier (with height $\Phi_B$) present at the LCMO/Ag interface. The presence of OV, which can electromigrate with the application of external voltage, locally determines the LCMO resistivity and also controls the height of the barrier. An interface with more (less) OV gives a high (low) resistance state. The barrier is not spatially uniform so the formation probability of conducting paths would be higher in those zones where the barrier is lower, as assumed in the model developed in Ref. \cite{Miranda_2017}. 

In the case of the NSTO (n-type)/LSMCO (p-type) device, the LSMCO perovskite is a redox topotactic system able to switch between oxidized and reduced perovskite phases with large differences in their number of carriers and conductivities (OV are electron donors). Upon electrical cycling, the redox reaction takes place, oxygen is exchanged between LSMCO and the ambient and this (reversibly) switches the depletion layer of the n-p diode, formed at the NSTO/LSMCO interface, by changing the amount of p carriers and thus the balance between n and p carriers at both sides of the junction. Oxidized LSMCO is linked to a thinner depletion layer and a low resistance state, while reduced LSMCO is related to a thicker depletion layer and a high resistance state \cite{Acevedo_2020,roman_2022}. The sketches displayed in Figures \ref{Fig1}(g), (h), (i) depict the mentioned mechanisms.

Figures \ref{Fig1}(d), (e), (f) displays the P-D response of the three systems. The P (D) curve is obtained after the application of successive SET (RESET) pulses. For the case of the LCMO-Ag device, P (D) was achieved by the application of +1.5V (-1.5V) pulses, with a time width of $\approx$ 1 ms. The conductance changes in a window between 0.79 $\mu$S and 0.54 mS, with 61 (114) conductance levels for P (D). In the case of the LCMO-Ti device, P (D) was obtained after the application of -2V (+2V) pulses, respectively. The time width of each pulses was also $\approx$ 1 ms. The conductance window in this case is between 15.9 mS and 55.5 mS. The number of levels were 35 and 101 for P and D, respectively. Finally, the P-D response corresponding to the NSTO-LSMCO device shows a conductance changing in the range 0.86 mS and 3.5 mS with 55 levels for P and 22 levels for D. For this, -2.5V and +5.5V pulses ($\approx$ 1 ms wide) were applied, respectively. The comparison of the three P-D responses shows differences in their conductance windows and number of levels. In addition, the three cases display non-linear and asymmetric P-D responses. In the next sections we will address how these differences in the P-D behaviour affect the convergence of simulated perceptrons.

\begin{figure}
  \centering    
  \includegraphics[scale=0.3]{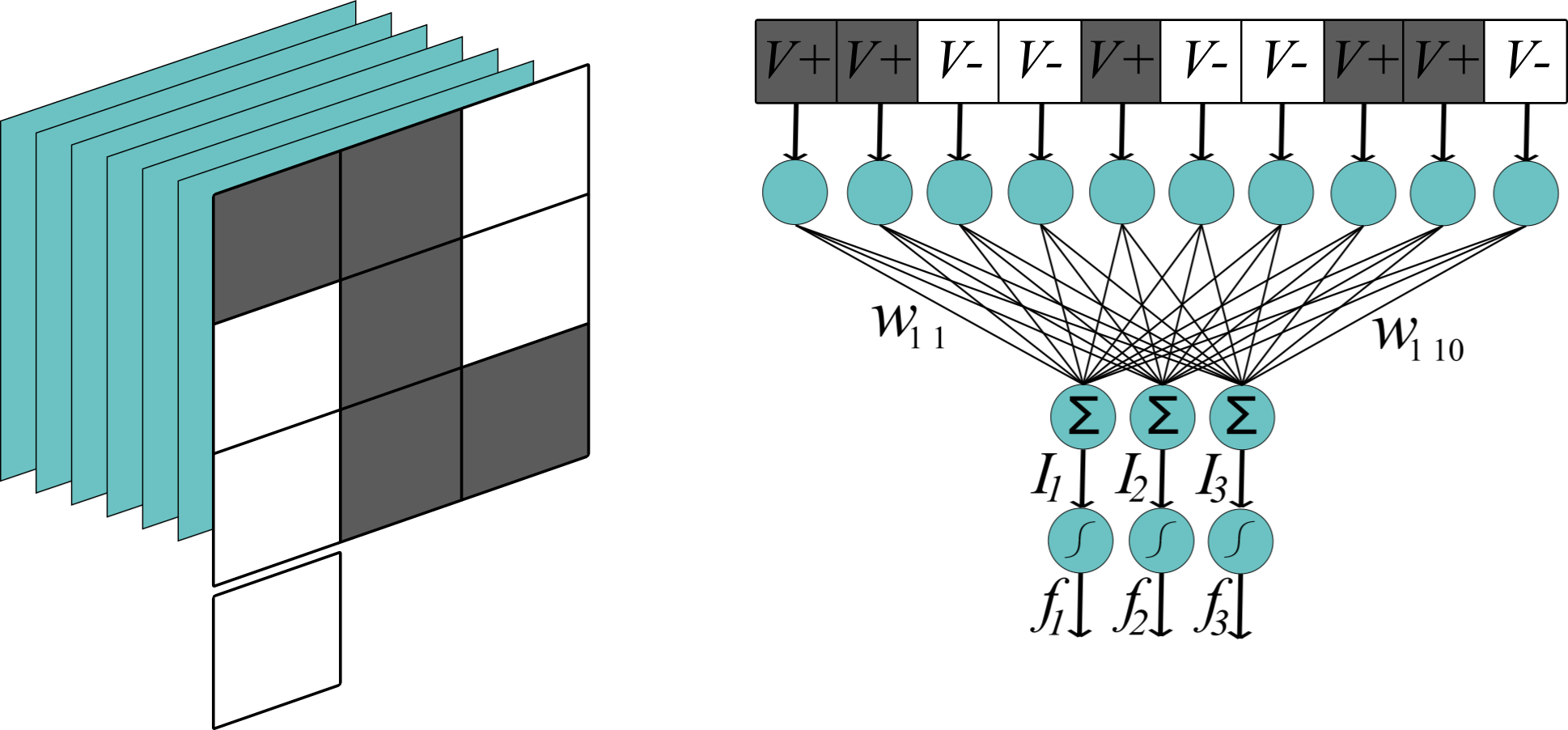}
  \caption{\textit{(Left) Example of a "\text{z}" character belonging to the used data-set used for the perceptron training. Each bright (dark) pixel is codified in voltage $V_{-} (V_{+})$ and  feeds the perceptron inputs; (Right) Sketch illustrating the simple perceptron we have used to simulate the training process for character recognition. See the main text for details}}
\label{Fig2}
\end{figure}

\section{Definition of the NN}

We have simulated a simple perceptron for character recognition, as displayed in the sketch of Fig. \ref{Fig2}. The NN is intended to classify three classes of characters, namely the letters "n" , "v" and "z", as implemented previously by Prezioso et al. \cite{Prezioso_2014}. Each character is represented by an image of 3x3 pixels, where each pixel could be bright or dark. Each class consists on 10 images obtained by flipping one pixel of the correct character. Each image is vectorized and feeds with voltages $V_j$ (-100 mV (+100 mV) for bright (dark) pixels) 9 bit lines of the NN (an additional bit line with a constant bias -100 mV is used).

The synaptic weights $w_{ij}$ of the NN are defined as a differential pair of memristors with conductances $G_{ij}^+$ and $G_{ij}^-$ of a 10x6 memristors cross-bar array, according to:

\begin{equation}
    w_{ij} = G_{ij}^+ - G_{ij}^-. 
\label{eq_weigth_map}
\end{equation}

This strategy allows obtaining either positive or negative synaptic weights from a physical magnitude -the conductance- which is  positive defined \cite{Alibart_2013}. The NN has three output lines, corresponding to the three classes of characters.  When an image $\mu$ feeds the NN, the output current of line $i$, following Eq.(1), is given by 

\begin{equation}
    I_{i,\mu} = \sum_{j=1}^{M} w_{ij} V_{j,\mu}.
\label{eq_current}
\end{equation}

An activation function,  

\begin{equation}
    f_{i,\mu} = \tanh(\beta I_{i,\mu}),
\label{eq_activation_func}
\end{equation}

is applied to each output line, with $\beta$ being a constant. We define as Loss function (L) the mean square error given by:

\begin{equation}
 L [w] = \frac{1}{2} \sum_{i \mu} \left( f_{i,\mu}^{(g)} - f_{i,\mu}  \right) ^2 \, ,
\label{eq_loss}
\end{equation}

where $f_{i,\mu}^{(g)}$ is +0.85 if an image $\mu$ corresponds to the class of the output $i$, and -0.85 otherwise (a similar approach was followed in Ref. \cite{Prezioso_2014}). 

The synaptic weight actualization $\Delta w_{ij}$ is calculated with the standard batch gradient descent method, according to:


\begin{equation}
    \Delta w_{ij} = - \eta \frac{\partial L}{\partial w_{ij} }= \eta \sum_{\mu} \Delta_{ij,\mu} ,
\end{equation}

where $\eta$ is the learning rate and $\Delta_{ij,\mu}$ is given by:

\begin{equation}
    \Delta_{ij,\mu} =V_{j\mu} [f_{i,\mu}^{(g)}-f_{i,\mu}]\frac{df_{i,\mu}}{dI}|_{(I=I_{i,\mu})} \; .
\end{equation}

In order to simplify the weight actualization process we have used the so-called Manhattan rule \cite{Zamanidoost_2015}, which only retains the sign of the exact actualization and applies constant SET (RESET) pulses (aimed to increase (decrease) the conductance of a specific memristor) according to the following rule: i) if $\Delta w_{ij} > 0$, then a SET (RESET) pulse to modify the conductance $G_{ij}^+$ ($G_{ij}^-$) is applied; ii) if $\Delta w_{ij} \leq 0$, then a RESET (SET) pulse is applied to modify the conductance $G_{ij}^+$ ($G_{ij}^-$).
This strategy is much easier to implement in a physical NN, as it avoids the determination and application of multiple write pulses to set each memristor in the target value, as it is usually needed for systems with non-linear P-D response. The evolution of the conductances $G_{ij}^+$ and $G_{ij}^-$ upon the application o SET/RESET pulses follow the P-D curves presented in the previous section.

Given the simplicity of the used dataset, which hampers the possibility of splitting it in train and test sets, we have used k-folds as a cross-validation method for the NN training \cite{Geron}. The convergence of the algorithm was determined by the standard procedure of tracking the evolution of the Loss function with the number of EPOCHs (we define for that the parameter EPOCHs-to-convergence, or ETC). We established a quantitative convergence criterion given by the absolute value of the derivative of the normalized Loss function with respect to the EPOCH number reaching a value $\leq 1\times 10^{-4}$. 
\section{Results}
\begin{figure}
  \centering    
\includegraphics[scale=0.47]{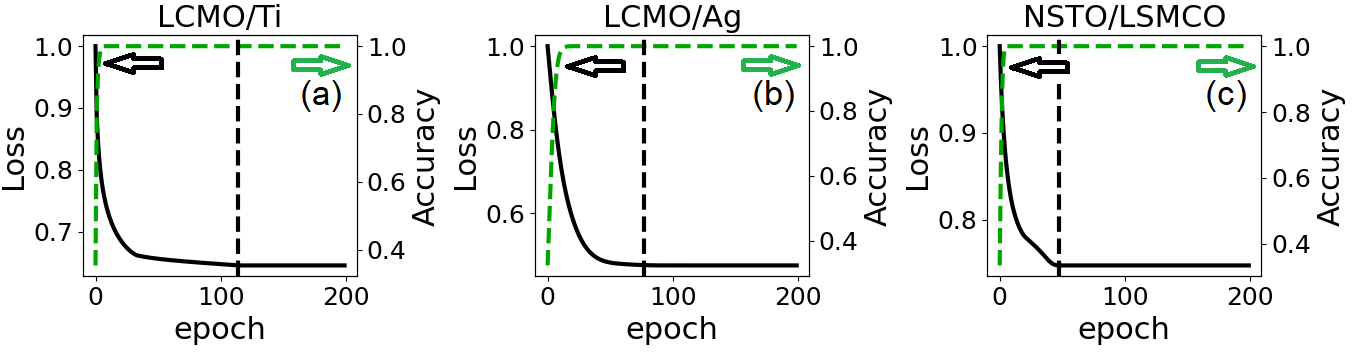}
\caption{\textit{Simulations displaying the evolution of the normalized Loss (solid black line) and the accuracy (dashed green line) during the learning process. P-D curves corresponding to the experimental (a) LCMO/Ti (b) LCMO/Ag and (c) NSTO/LSMCO memristors were used. Black dashed lines indicate the EPOCHs-to-convergence (ETC) in each case.}}
\label{Fig3}
\end{figure}

Fig. \ref{Fig3} shows simulations with the evolution of the Loss as a function of the number of EPOCHs for the three experimental systems, with P-D responses already shown in Fig. \ref{Fig1}. 2000 realizations were made for each system, where in each realization the conductances $G_{ij}^+$ and $G_{ij}^-$ were randomly chosen in their allowed conductance bands, and the displayed curves are the average between all the realizations for each case. It is seen that LCMO-Ag, LCMO-Ti and NSTO-LSMCO systems converge after 77, 114 and 47 EPOCHs, respectively. The same figure displays the evolution of the accuracy (defined as the ratio between well classified patterns against the total number of patterns) with the number of EPOCHs, which is found to achieve a value of 1 in fewer EPOCHs than the ETCs extracted from the evolution of the Loss. This difference relies on the simplicity of the used data set -reported as linearly separable for two classes problems \cite{Alibart_2013}- and indicates that the evolution of the Loss is a  subtler indicator to track the robustness of the convergence, as it is sensitive to the achievement of optimal distances between the decision boundaries and the data set.

We notice that  is difficult to explain the different ETCs observed for LCMO-Ag, LCMO-Ti and NSTO-LSMCO cases, as their P-D responses differ in more than one feature (range of conductance, number of levels, non-linearity/asymmetry). In order to address separately the influence of each one in the ETCs response, we have run new simulations using synthetic P-D curves that allow to vary these features in a one-by-one basis.

\begin{figure}
  \centering    
  \includegraphics[scale=0.75]{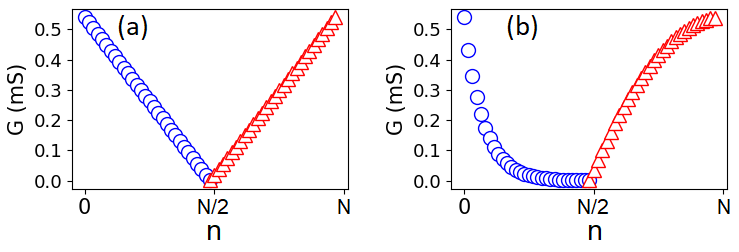}
  \caption{\textit{Linear/symmetric (a) and non-linear/asymmetric (b) synthetic potentiation (P, red symbols) and depression (D, blue symbols) curves used to simulate the perceptron learning process. We have used different values of N ranging from 12 to 2000. See the main text for further details.}}
\label{Fig4}
\end{figure}

Figure \ref{Fig4} shows a sketch of the synthetic P-D curves. We have fixed the conductivity range between 0.79 $\mu$S and 0.54 mS. We have simulated linear/symmetric (panel (a)) and non-linear/asymmetric (panel (b)) curves, the latter following laws given by $G = \alpha e^{-\gamma n} $ (D) and $G = -an^2 + bn + c$ (P), where n is the number of pulses and $\alpha$, $\gamma$, a, b and c are constants. We chose these fitting functions to resemble the behavior of the LCMO-Ti P-D curves (Figure \ref{Fig1}(a)), including the observed asymmetry. For both cases the number of levels (N) was varied between 12 and 2000. We notice that real memristive systems usually can display P-D curves with up to $\approx$ 10$^2$ levels; however, it is worth simulating higher N values to determine the evolution of the network behavior when the P-D curves granularity increases and approaches a quasi-continuum situation.

\begin{figure}
  \centering    
  \includegraphics[scale=0.7]{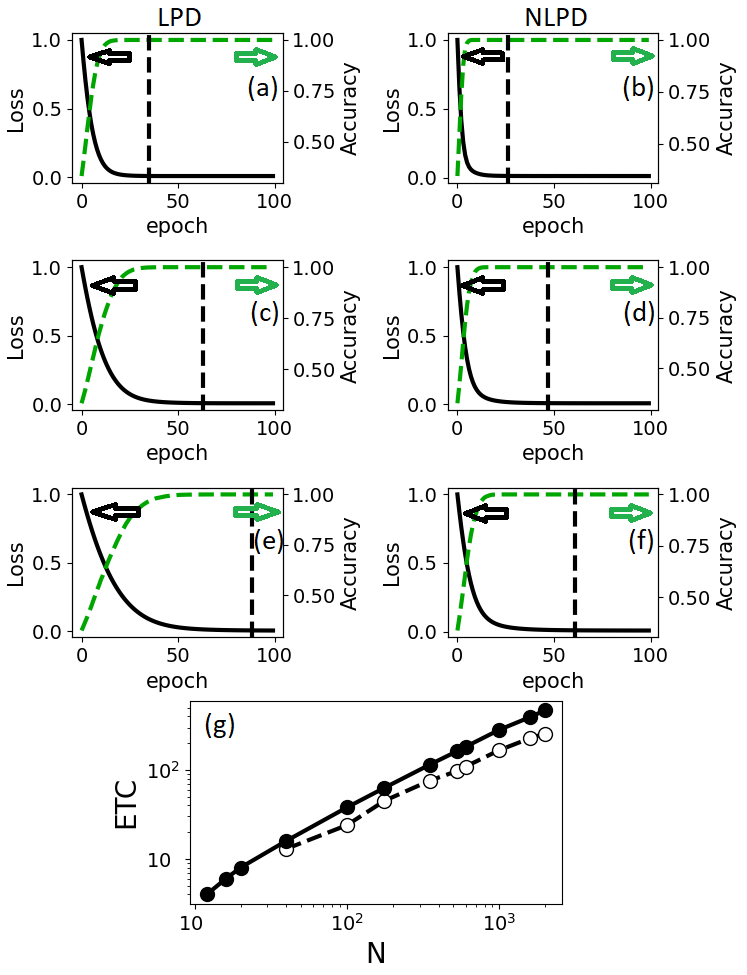}
  \caption{\textit{Evolution of the normalized Loss (solid black line) and the accuracy (green dashed line) for the learning process using synthetic linear/symmetric and non-linear/asymmetric P and D curves with different number of levels N. (a) linear case with N =88, (b) non-linear case with N = 88; (c) linear case with N =175, (d) non-linear case with N = 175; (e) linear case with N = 263; (f) non-linear case with N = 263. Black dashed lines indicate the EPOCHs to convergence (ETC) in each case. Panel (g) displays the ETC as a function of N for linear (black symbols) and non-linear (white symbols) P-D curves.}}
\label{Fig5}
\end{figure}

Figure \ref{Fig5} shows the evolution of the Loss as a function of the number of EPOCHs for linear/symmetric and non-linear/asymmetric synthetic P-D curves and N = 88, 175 and 263, respectively. The evolution of the accuracy is also shown, for completeness. Again, the displayed curves are the average over 2000 realizations. Several features, summarized in Fig. \ref{Fig5} (g), are observed: i) the ETC is seen to decrease as N decreases. This is an expected result related to the Manhattan rule used for the weight actualization, as a lower N implies a lower granularity for the P-D curves, which impacts on larger actualization values $\Delta w_{ij}$ that can be interpreted as a higher \emph{effective} learning rate (we recall that no formal learning rate is used with the Manhattan rule); ii) no convergence -reflected in an oscillatory behavior of the Loss with the EPOCHs- was achieved for N < 12 (linear P-D curve) and N < 40 (non-linear P-D curve). In this situation, the \emph{effective} learning rate is too high to allow the stabilization of the Loss in a minimum; iii) the ETC of non-linear/asymmetric P-D curves (white circles) is lower that the one obtained for linear/symmetric P-D curves (black circles), for all values of N where convergence was achieved for both P-D curves. This might be (again) related to the Manhattan rule used for the weight actualization and to the non-linearity of the P-D curves, as those weights displaying relatively low (high) conductances that need to be potentiated (depressed) during the training are updated with higher values of $\Delta w_{ij}$ for the first EPOCHs, which are progressively reduced afterwards. This process resembles some of the strategies used to optimize the convergence of software based NN that use an \emph{adaptive effective} learning rate  (i.e., in Keras).

 We notice that an improved convergence for non-linear/asymmetric P-D curves is at odds with the usual assumption that a linear/symmetric P-D is more appropriate to optimize the learning process of a physical NN \cite{Burr_2015,Chen_2015,Shibata_2020,Apsangi_2022}, but is in agreement with other reports that indicate that this issue might be dependent on the weight actualization method \cite{Gutsche_2021}.

Figs. S1 and S2 of the Supplementary Information show additional simulations where the conductances range was varied for fixed N. It is found that the ETC decreases as the conductivity range increases, both for linear and non-linear P-D curves. Again, this indicates that a higher conductance window leads into higher \emph{effective} learning rates that speed the convergence. In summary, our results show that, for the chosen weight actualization (the Manhattan rule), the ETC can be optimized for non-linear/asymmetric P-D curves with relatively low granularity.

\begin{figure}
  \centering    
  \includegraphics[scale=0.55]{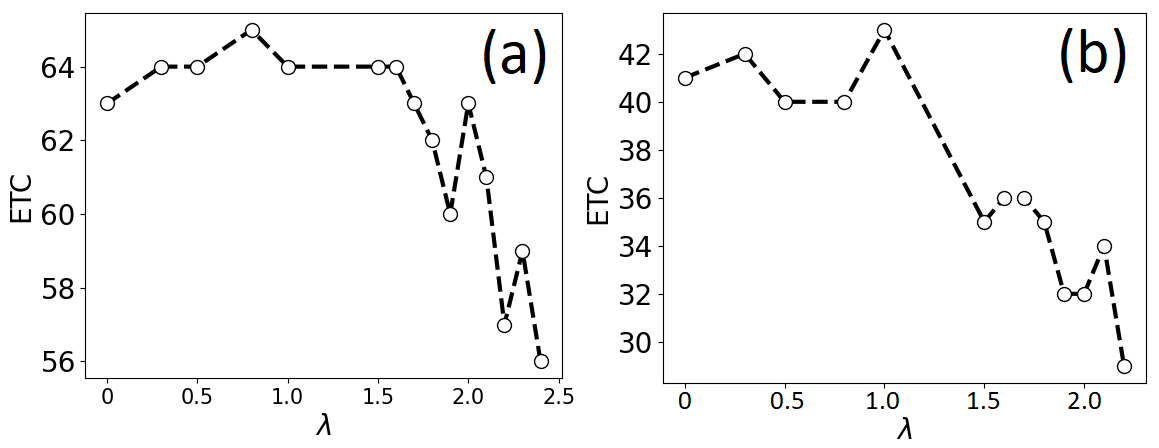}
  \caption{\textit{Evolution of the EPOCHs to convergence (ETC) vs. $\lambda$ (which controls the intensity of the injected noise during the synaptic weight actualization) for linear/symmetric (a) and non-linear/asymmetric (b) synthetic P-D curves. See the main text for further details.}}
\label{Fig7}
\end{figure}

Finally, we demonstrate that the ETC can be reduced by adding noise to the weight actualization. We notice that, as we mentioned earlier, the introduction of noise is a well reported strategy to improve the convergence of standard neural networks; however, its feasibility for a physical network where the weights are updated by the Manhattan rule needs to be tested. To this end we run simulations where the synaptic weights were updated according to:

\begin{equation}
    \Delta w'_{ij} = \Delta w_{ij} (1 + p \lambda),
    \end{equation}

where $\Delta w_{ij}$ is the weight actualization given by the Manhattan rule, $p$ is a number  randomly chosen in the interval [-1,1] from an uniform distribution in each actualization instance and $\lambda$ is a parameter that controls the intensity of the injected noise. Fig. \ref{Fig7} shows the evolution of the ETC as a function of $\lambda$ for systems trained with noise injection, both for linear/symmetric and non-linear/asymmetric P-D responses. The conductance range was set between 0.79 $\mu$S and 0.54 mS and the number of levels of the P-D curves was fixed in N=175. It is seen that for the linear/symmetric P-D curves the injection of noise slightly decreases the ETC from 63 EPOCHs (without noise) to 56 EPOCHs (for $\lambda$ = 2.4), that is  a reduction of $\approx$ 11 \%. However, in the case of the non-linear/asymmetric P-D curves, the ETC is reduced more significantly, from 41 EPOCHs (without noise) to 29 EPOCHs (for $\lambda$ = 2.2), that is $\approx$ 29 \%. We notice that for $\lambda$ > 2.4 and $\lambda$> 2.2 the NN no longer converges for linear/symmetric and non-linear/asymmetric P-D curves, respectively. These results show that noise injection is a suitable strategy to optimize the Loss convergence to a global minimum, avoiding its stabilization in local minima and thus the delay of the learning process. 
In addition from the comparison of Figs. \ref{Fig7} (a) and (b) it can be conclude that for  a given value of $\lambda$, the reduction in the ETC is higher for the non-linear/asymmetric P-D curves.

\section{Conclusions}

In summary, we have studied the influence of specific characteristics of the P-D curves  such as granularity and nonlinearity/asymmetry on the convergence of simple perceptrons based on memristor cross-bar arrays. Our simulations show that the fastest convergence (minimum ETC) is achieved for non-linear/asymmetric P-D curves with low granularity, always in a framework of weight actualization according to the Manhattan rule. Further, we have shown that the injection of noise on the weight actualization substantially improves the ETC in the case of non-linear/asymmetric P-D curves. This might be relevant for NN that need continuous retraining. We also notice that these results might be dependent on the weight actualization method and deserve further studies to determine their validity when other protocols are used. Also, validation by using more complex data-sets (i.e. the MNIST database), where a trade-off between ETC and the accuracy might be anticipated, should be performed and will be addressed in the future.

\vspace{10px}

\textbf{Acknowledgements}

We acknowledge support from ANPCyT (projects PICT2019-02781, PICT2019-00654 and PICT2020A-00415) and EU-H2020-RISE action "MELON" (Grant No. 872631). We thank W. Román Acevedo for the electrical measurements.

\vspace{10px}

\textbf{Supplementary Information}

See Suplementary Information for additional simulations.

\vspace{10px}

\textbf{Conflict of Interest}

The authors have no conflicts to disclose.

\vspace{10px}

\textbf{Data availability}

The data that support the findings of this study are available from the corresponding author upon reasonable request.

\bibliography{references}

\end{document}